\documentclass[aip,amsmath,reprint,twocolumn]{revtex4-1}
\usepackage[usenames]{color}
\usepackage{ae,aecompl}
\usepackage[normalem]{ulem}
\usepackage{graphicx}
\usepackage{dcolumn}
\usepackage{bm}
\begin{document}

\title{Controlling Tokamak Geometry with 3D Magnetic Perturbations}

\author{T M Bird}
\affiliation{Max Planck Institute for Plasma Physics, EURATOM Association, Wendelsteinstr. 1, 17491 Greifswald, Germany}
\author{C C Hegna}
\affiliation{Departments of Engineering Physics and Physics, University of Wisconsin-Madison, 1500 Engineering Dr., Madison WI 53703, USA}
\email{tbird@ipp.mpg.de}

\begin{abstract}
It is shown that small externally applied magnetic perturbations can significantly alter important geometric properties of magnetic flux surfaces in tokamaks.  Through 3D shaping, experimentally relevant perturbation levels are large enough to influence turbulent transport and MHD stability in the pedestal region.  It is shown that the dominant pitch-resonant flux surface deformations are primarily induced by non-resonant 3D fields, particularly in the presence of significant axisymmetric shaping.  The spectral content of the applied 3D field can be used to control these effects.
\end{abstract}
\maketitle

Very small externally applied non-axisymmetric magnetic perturbations can significantly alter the behavior of the edge plasma in tokamaks\cite{lang_nf13}.  Significant changes in transport and/or macroscopic stability have been observed in most of the world's major tokamak experiments when these perturbations are applied \cite{evans_prl04, evans_nf08, suttrop_prl11, liang_prl07, nardon_jnm09, kirk_nf10, kstar_1}, and a coil set to apply these perturbations is now included in the design of the next-step ITER experiment.  The helicity of these perturbations is often chosen to be pitch resonant with magnetic field lines in the edge, hence they are usually referred to as Resonant Magnetic Perturbations (RMPs).  The enormous gradients in temperature and density which arise in the tokamak edge during High confinement mode (H-mode) operation are essential for achieving reactor-relevant performance, but these gradients provide an equally enormous source of free energy for explosive macroscopic instabilities called Edge Localized Modes (ELMs) that periodically expel hot plasma onto the device walls.  The heat loads to plasma facing components are tolerable in existing machines \cite{leonard_ppcf06}, but are anticipated to be problematic in reactor-scale experiments such as ITER \cite{loarte_ppcf03, loarte_iaea10}.  The prospect of taming ELM-induced heat loads to the wall using RMPs is enticing, hence a great deal of effort is now being expended to understand and exploit their effect.  

The effect of RMPs on Tokamak plasmas is remarkable and puzzling because of their small magnitude - 3D fields whose strength is $10^{-3}-10^{-4}$ of the background toroidal field strength are now routinely used to modify ELM behavior in existing Tokamaks.  It was initially thought that RMPs would reduce the confinement properties in the edge region via stochastic transport associated with induced overlapping magnetic islands.  The resulting enhanced transport would reduce the free-energy source for ELM-inducing instabilities \cite{evans_nature}.  However, toroidally rotating plasmas respond to externally produced resonant magnetic perturbations by driving screening currents at the associated rational surface \cite{fitzpatrick_nf93}.  These currents can partially or completely cancel the topology-breaking perturbation, maintaining the integrity of flux surfaces.  The most recent simulations with extended MHD codes suggest that this effect is operative in most experimental scenarios, and that the RMPs can most likely only significantly alter the magnetic topology in radially localized regions where the electron rotation passes through zero \cite{ferraro_pop12, becoulet_nf12, liu_nf11}.  Recent experimental measurements across a number of different machines suggest that flux surfaces remain largely intact in the plasma edge and that they often develop 3D deformations on the order of $1-5 \%$ of the minor radius \cite{itpa_mhd}.  The intensity of turbulent density fluctuations during RMP experiments at DIII-D have been directly measured, and they exhibit a direct and complicated sensitivity to the applied 3D fields \cite{mckee}.  In this work we provide a theoretical explanation for how these external perturbations affect the 3D shaping that subsequently alters pedestal stability and turbulent transport.

Our focus is on how 3D magnetic fields deform magnetic flux surfaces, and the consequences for stability and transport.  The safety factor is typically $\geq 3$ in the pedestal and thus resonant helical fields typically have a large poloidal mode number in the region of interest.  The fine scale structure of these perturbations means that the resulting displacement of flux surfaces has a surprisingly potent effect on the magnetic curvature and local magnetic shear.  The magnetic curvature is determined mostly by gradients in the magnetic field strength, so a small perturbation can induce significant curvature if it oscillates rapidly in space.  Experimentally relevant magnetic perturbations are shown to dramatically modify infinite-n ideal MHD ballooning stability, which has been found to be a good proxy for the onset of Kinetic Ballooning Mode driven microturbulence which is thought to limit the achievable (local) pressure gradient inside the pedestal \cite{eped1}.  Recent work has also shown that 3D fields can also adversely affect intermediate-n peeling-ballooning stability \cite{hegnapop}.

The relationship between the 3D magnetic field spectrum and the spectrum of deformations of a flux surface is determined by a geometric coupling matrix which is a complicated function of aspect ratio, 2D shaping, the spacing between flux surfaces, and the pitch-alignment between the perturbations and the equilibrium magnetic field lines.  We show that the non-resonant components of the 3D field spectrum play the most important role in deforming the shape of flux surfaces.  This may be able to explain many strange results of RMP experiments, as most of the focus in the community so far has been mostly on the resonant perturbations.

In order to make analytic progress a number of simplifications are used.  Primarily we restrict ourselves to situations where magnetostatic force balance is maintained in regions with topologically toroidal magnetic flux surfaces.  This should occur when plasma screening is sufficiently strong to maintain the integrity of flux surfaces.  Here, what we are interested in describing is the 3D distortion of the flux surface shape.  In principle, the results presented here should be applicable even if physics beyond ideal MHD plays a role in determining the 3D fields in the plasma.  

In order to describe the geometry of the perturbed flux surface, we utilize local equilibrium theory \cite{miller, local3D_1, local3D_2}.  This technique allows one to construct solutions to the ideal MHD equilibrium equations in the vicinity of a particular magnetic flux surface.  In this formulation, the position of the flux surface is expressed using an inverse coordinate transformation using two straight field line angles $\theta$ and $\zeta$.  In cylindrical coordinates, $\vec{x}_{0}(\theta,\zeta) = [R(\theta,\zeta) \hat{R}, \phi(\theta,\zeta) \hat{\phi}, Z(\theta,\zeta) \hat{Z}]$.  In the vicinity of the flux surface of interest, a Taylor expansion in the toroidal flux surface label $\psi$ is given by $\vec{x}(\psi,\theta,\zeta)= \vec{x}_{0}(\theta,\zeta) + (\psi-\psi_{0}) \partial \vec{x}(\psi,\theta,\zeta) / \partial \psi + O((\psi-\psi_{0})^{2})$.  Specification of two profile quantities at the surface and solutions to the MHD equilibrium equations constrain the value of $\partial \vec{x} / \partial \psi$ and determines the Jacobian for the transformation, $\sqrt{g} = \partial \vec{x} / \partial \psi \cdot \partial \vec{x} / \partial \theta \times \partial \vec{x} / \partial \zeta$.  

In axisymmetry, this procedure is equivalent to the Taylor expansion in ($\psi - \psi_{0}$) of the Grad-Shafranov equation.  If the symmetry angle is taken to be the straight field line toroidal angle $\zeta$, the Jacobian satisfies $\sqrt{g} = R^{2} V' / < R^{2} >$ where the bracket denotes a flux surface average and $V' = dV / d \psi$ provides an overall normalisation factor where V is the volume enclosed by a flux surface.  For non-axisymmetric plasmas where Grad-Shafranov theory is not applicable, the geometric constraint that no currents flow normal to the flux surface provides a condition to determine $\sqrt{g}$ and ensure that the ideal MHD equilibrium equations are satisfied locally \cite{local3D_2}.  This condition takes the form of a first order partial differential equation for $\sqrt{g}$ given by
\begin{equation}
\frac{\partial}{\partial \theta} \frac{ q g_{\zeta \zeta} + g_{\zeta \theta} }{\sqrt{g}} = \frac{\partial}{\partial \zeta} \frac{ q g_{\theta \zeta} + g_{\theta \theta}}{\sqrt{g}}
\end{equation}
where q is the safety factor at the surface of interest, and $g_{\theta \theta} = \frac{\partial \vec{x}_{0}}{ \partial \theta} \cdot \frac{\partial \vec{x}_{0}}{ \partial \theta}, g_{\theta \zeta} =  \frac{\partial \vec{x}_{0}}{ \partial \theta} \cdot \frac{\partial \vec{x}_{0}}{ \partial \zeta}, g_{\zeta \zeta} =  \frac{\partial \vec{x}_{0}}{ \partial \zeta} \cdot \frac{\partial \vec{x}_{0}}{ \partial \zeta }$ are metric elements of the coordinate transformation.  In general this equation must be solved numerically, and a lightweight code has been written to do this (the 3D Local Equilibrium (3DLEQ) code, used in \cite{birdhegna}).  With the jacobian in hand, the equilibrium is determined, and one can then calculate all of the MHD equilibrium quantities - for example $\vec{B} = \frac{1}{\sqrt{g}}( \frac{\partial \vec{x}_{0}}{\partial \zeta} + \frac{1}{q}\frac{\vec{x}_{0}}{\partial \theta}).$  The unit vectors $\hat{b} = \vec{B} / |B| , \hat{n} = \vec{\nabla \psi} / | \nabla \psi|$ can be calculated from $\vec{x}_{0}$ and from these one can construct a Frenet-Serret frame which succinctly describes the geometry of magnetic field lines.  For example, the curvature vector is simply $ \vec{\kappa} = \hat{b} \cdot \nabla \hat{b} = \kappa_{n} \hat{n} + \kappa_{g} \hat{b} \times \hat{n}$ with $\kappa_{n}$ the normal curvature and $\kappa_{g}$ the geodesic curvature.  The usefulness of local equilibrium theory is that one can calculate the curvature quantities exactly as a function of the flux surface parametrization.

With a description of the equilibrium one can then calculate how the 3D fields deform flux surfaces, and how this affects quantities like the magnetic curvature.  We now apply this procedure to analytic solutions for a simple equilibrium, which yield some surprising insights about the role of small 3D deformations to the shape of magnetic flux surfaces.  A high aspect ratio, circular cross section Tokamak equilibrium with a small, single helicity 3D deformation is considered.  The position of the flux surface of interest in cylindrical coordinates ($\vec{x}_{0} = [ R, \phi , Z]$) is given by
\begin{eqnarray}\label{parametrization1}
R = R_{0} + \rho cos ( \theta ) + \gamma cos ( \alpha ) , \\ 
Z = \rho sin ( \theta ) + \gamma sin ( \alpha ) ,
\end{eqnarray}
and $\phi = -\zeta$.  Here, $\theta$ is the geometric poloidal angle, $\zeta$ is the geometric toroidal angle, and $\alpha = m\theta-n\zeta $ is a helical angle determined by the helicity of the deformation.  For the high aspect ratio case here with no 2D shaping, $\theta$ is to lowest order a straight field line angle. A perturbation approach is now used to obtain approximate formulae for key geometric quantities.  We choose the ordering $\rho / R_{0} \sim \gamma / \rho \sim 1 / m \sim 1 / n^{2} \sim 1 / q^{2} \sim |m/q-n|^{2} \sim \epsilon \ll 1$ and then solve the local equilibrium equations analytically at each order in $\epsilon$.  The ordering is motivated by the properties of measured 3D deformations in the pedestal region during typical DIII-D RMP experiments, which are pitch-resonant with $\gamma / \rho \sim 10^{-2}$, $m \geq 10$, $n=3$, $q \geq 3$. To O($\epsilon^{1}$), the jacobian and field strength are unaffected by the 3D perturbation and are simply given by their values in the axisymmetric limit, $\sqrt{g} = V' ( 1 + 2 \rho cos (\theta) / R_{0} )$ and $|B| = B_{0} / (1 + \rho cos (\theta) / R_{0} )$ where $B_{0}$ is determined by the normalisation factor $V' = R_{0} / B_{0} $.  However, the 3D effects enter at lowest order for a number of quantities.  The lowest order $| \nabla \psi|$ is 
\begin{equation}
|\nabla \psi| = \frac{B_{0} L_{\psi}^{2}}{R_{0}}
\end{equation}
where $L_{\psi}^{2} = R_{0} \sqrt{ \rho^{2} + \gamma^{2} m^{2} + 2 \rho \gamma m cos ( \alpha_{-1} ) }$ is the scale factor for the radial coordinate and $\alpha_{-1} = (m-1) \theta - n \zeta$.  We can see that the ratio $\gamma m / \rho$ provides a rough estimate for the contribution of high-m 3D deformations.  High-m deformations with $\gamma / \rho \sim 10^{-2}$ can thus modulate the poloidal field strength ($\vec{B} \cdot \nabla \theta = | \nabla \psi| / R$) in the edge region.  The result is that a fine scale deformation of the position of the flux surface can modulate the direction in which the normal vector points, even if the actual position of the surface is only modulated slightly.  This has serious consequences, because the stability of a plasma is very sensitive to these geometric details.  The lowest order normal and geodesic curvatures (determined by the $O(\epsilon)$ equilibrium equations) are now given by
\begin{eqnarray}\label{curvatures}
\kappa_{n} = \frac{ -\rho cos ( \theta ) - \gamma m cos ( \alpha )}{L_{\psi}^{2} }, \kappa_{g} = \frac{ \rho sin ( \theta ) + \gamma m sin ( \alpha )} {L_{\psi}^{2}}.
\end{eqnarray}
These calculations suggest that the curvature produced by small fine-scale perturbations can compete with the toroidal curvature.  The magnitude and helicity of flux surface displacements measured in many experiments suggest that the magnitude of the 3D terms in these expressions will be significant and in some cases perhaps even dominant \cite{chapman_nf07, chapman_nf12, moyer_nf12, ferraro_nf13, itpa_mhd}.   

In a sheared magnetic field, the most virulent instabilities tilt in the plane perpendicular to $\vec{B}$ as they propagate along field lines.  The tilting of these instabilities is determined by the field-line-integrated local magnetic shear ($\Lambda = \frac{|\nabla \psi|^{2}}{B} \int \frac{B^{2}}{|\nabla \psi|^{2}} \sqrt{g} s  d\zeta$ where $s = (\hat{b} \times \hat{n}) \cdot \nabla \times (\hat{b} \times \hat{n})$ is the local magnetic shear).  This tilting effectively modulates the region of instability drive: it is not simply the region where $\kappa_{n}$ is negative, but the region where $\kappa_{n} + \Lambda \kappa_{g}$ is negative.  This geometric quantity determines the drive terms for pressure/temperature gradient driven local instabilities in the context of ballooning formalism.  

To highlight what physical mechanisms determine the local magnetic shear, we use the identity $s = \mu_{0} \vec{J} \cdot \vec{B} / B^{2} - 2 \tau_{n} $ \cite{local3D_2}.  Previously, it has been shown that parallel currents can play an important role.  In particular, near rational surfaces, 3D deformations can trigger near-resonant Pfirsch-Schluter currents that substantially modulate the local magnetic shear \cite{birdhegna}.  The second term in this equation is the 'normal torsion', which can be defined as $\tau_{n} = -\hat{n} \cdot ( \hat{b} \cdot \nabla) \hat{b} \times \hat{n}.$  The normal torsion is even more sensitive to fine scale deformations than the curvature quantites.  For the high-m perturbed equilibrium considered here, the lowest order torsion is
\begin{eqnarray}\label{taun}
\tau_{n} = \frac{1}{L_{\psi}^{4}} \Bigg[ R_{0} ( m/q - n ) \gamma m ( \rho cos \alpha_{-1} + \gamma m ) + \nonumber\\
 \frac{R_{0} \rho}{q} ( \rho + \gamma m cos_{\alpha_{-1}} ) 
+ 3 \rho cos \theta \left (\frac{\gamma m^{2} \rho}{q} cos \alpha_{-1} + \frac{\gamma^{2} m^{3}}{q} \right ) \Bigg]
\end{eqnarray}
Very small ($dR / \rho \sim 10^{-2}$) 3D deformations can thus substantially modulate a number of important geometric quantities that influence pressure/curvature driven instabilities.  

\begin{figure}
\centering
\includegraphics[width=6.5cm]{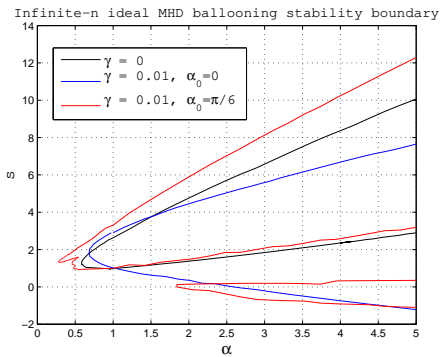}
\caption{The infinite-n ideal MHD ballooning stability boundary for circular cross section Tokamak equilibria with $\rho/R_{0} = 0.33$, $q=11/3 + 0.05$.  A 3D deformation with $\gamma / \rho = 0.03$, $m=11$, $n=3$ strongly perturbs the stability boundary.  This deformation is produced by a radial magnetic perturbation with magnitude $B_{\rho} / B_{0} = 2.6e-4$ and helicity $m/n = 10/3$, which is non-resonant. }
\label{3d_localshear}
\end{figure}

The destabilizing effect of fine scale non-axisymmetric perturbations is demonstrated in Figure \ref{3d_localshear}.  The infinite-n ballooning stability boundary is shown for a circular cross section equilibrium with a 3D deformation whose magnitude is $3 \%$ of the minor radius.  Stability diagrams for two different choices of the field line label $\alpha_{0} = q \theta - \zeta$ are shown - for each choice of $\alpha_{0}$ the radial wave-vector is varied along the field line to find the most unstable eigenvalue.  Both the equilibria and the stability diagrams are generated using the process described in Ref. \cite{birdhegna}.  All of the previously discussed mechanisms are active here.  The ballooning stability physics will be discussed in detail in a future publication but we present these calculations as numerical evidence to support our analytic calculations.  This demonstrates conclusively that ballooning stability is highly sensitive to experimentally relevant 3D deformations \cite{kirk, itpa_mhd, ferraro_nf13}.  

The relationship between the 3D magnetic field spectrum and the 3D deformation of flux surfaces has some surprising subtleties.  We can calculate the 'radial' (relative to the unperturbed 2D flux surface) magnetic perturbation associated with 3D deformations as $B_{\rho} = \vec{B}^{3D} \cdot \vec{  \nabla \psi }^{2D}$ where $\vec{B}^{3D}$ is the magnetic field calculated using only the 3D components of Equations 2-3 and $\vec{ \nabla \psi }^{2D}$ is calculated with the 3D deformations set to 0.  Let us now consider a flux surface with arbitrary 2D shape and a generalized spectrum of 3D deformations, such that
\begin{eqnarray}
R = R(\Theta) + \sum_{i} \gamma^{1}_{i} cos( \alpha_{i} ) - \gamma^{2}_{i} sin (\alpha_{i} ) \\
Z = Z(\Theta) + \sum_{i} \gamma^{1}_{i} sin( \alpha_{i} ) + \gamma^{2}_{i} cos (\alpha_{i} )
\end{eqnarray}
where $\alpha_{i} = m_{i} \Theta - n \zeta$ and $\Theta$ is a straight field line angle which is in general not equal to the geometric poloidal angle.  For small 3D perturbations, we will relate the ampltiudes of the 'radial' magnetic field's spectral components to the quantities $\gamma_{i}$ through a linear coupling matrix of the form $\vec{B_{\rho}} = \bar{\bar{A}} \vec{\gamma}$.

The axisymmetric $\vec{\nabla \psi}$ is given by $\vec{\nabla \psi}^{2D} = R^{2D} / \sqrt{g} \times ( Z_{\Theta}^{2D} \hat{R} - R_{\Theta}^{2D} \hat{Z} ) $.  We decompose the derivatives as $R_{\Theta}^{2D} (\Theta) = \sum_{k} r_{k}^{c} cos( k \Theta ) + r_{k}^{s} sin (k \Theta ) $ and $Z_{\Theta}^{2D} (\Theta) = \sum_{k} z_{k}^{c} cos( k \Theta ) + z_{k}^{s} sin ( k \Theta ) .$
The 3D magnetic field to leading order is given by 
\begin{eqnarray}
\vec{B}^{3D} = \frac{1}{V'} \sum_{i} (n - m_{i}/q ) [ \gamma^{1}_{i} sin ( \alpha_{i} ) + \gamma^{2}_{i} cos ( \alpha_{i} ) \hat{R} \nonumber\\
 - \gamma^{1}_{i} cos ( \alpha_{i} ) + \gamma^{2}_{i} sin ( \alpha_{i} ) \hat{Z} ] .
\end{eqnarray}
The pre-factor of $(n-m_{i}/q)$ is the manifestation of field line bending physics - a flux surface deformation which is pitch resonant with the equilibrium magnetic field lines can be induced by smaller magnetic fields (relative to non-resonant deformations) because they minimize the amount of field line bending required to deform the flux surfaces.  This may be able to explain the recent observation in global 3D MHD equilibrium calculations that the deformations near rational surfaces tend to be dominated by deformations with helicity $m/n$ and $(m \pm 1)/n$  \cite{turnbull_pop13}.  However, $B^{3D}$ has not yet been projected into the 'radial' direction to calculate $B_{\rho}$.  Projecting $B^{3D}$ into the axisymmetric $\vec{\nabla \psi}$ direction gives us
\begin{eqnarray}\label{fullcoupling}
B_{\rho} = \frac{R^{2D}}{\sqrt{g}} \frac{1}{2} \sum_{i, k} ( n - m_{i}/q ) [ C_{k} cos ( \alpha_{+k} ) 
+ D_{k} cos ( \alpha_{-k} )  \nonumber\\
+ E_{k} sin ( \alpha_{+k} ) 
+ F_{k} sin ( \alpha_{-k} ) ] .
\end{eqnarray}
where $C_{k} = ( \gamma^{1}_{i} r_{k}^{s} + \gamma^{2}_{i} z_{k}^{s} + \gamma^{2}_{i} r_{k}^{c} - \gamma^{1}_{i} z_{k}^{c} ) $, $D_{k} = ( \gamma^{2}_{i} r_{k}^{c} - \gamma^{1}_{i} z_{k}^{c} - \gamma^{1}_{i} r_{k}^{s} + \gamma^{2}_{i} z_{k}^{s} ) $, $E_{k} = (\gamma^{1}_{i} r_{k}^{c} + \gamma^{2}_{i} z_{k}^{c} + \gamma^{2}_{i} r_{k}^{s} - \gamma^{1}_{i} z_{k}^{s})$, $F_{k} = (\gamma^{1}_{i} r_{k}^{c} + \gamma^{2}_{i} z_{k}^{c} - \gamma^{2}_{i} r_{k}^{s} + \gamma^{1}_{i} z_{k}^{s} ) $ .

It's clear from this expression that 2D shaping is important here, which we see through the poloidal spectrum of $\nabla \psi$.  The kth harmonic of $R_{\Theta}$ and $Z_{\Theta}$ allows the radial magnetic field with helicity $(m_{i} \pm k)/n$ to couple to the deformation with helicity $m_{i}/n$.  The q profile and surface averaged shear play an important role as they determine the radial variation of the $(n - m/q)$ prefactor.  The radial variation of the 3D field spectrum is also important, though it is determined by global physics which is beyond the scope of our local analysis.  The 3D field spectrum from global equilibrium calculations can be used as the input for the local analysis we have utilized here.   

We return to the high aspect ratio, circular cross section equilibrium and consider a spectrum of 3 deformations labeled by $i=-1, 0, 1$ with poloidal mode numbers $m_{i} = m-1 , m, m+1$ respectively.  Here we use the ordering $\rho / R_{0} \sim \gamma / \rho \sim \epsilon \ll 1$ but make no assumption about the magnitude of $m_{i}$, n and q.  Now, $Z_{\theta} = \rho cos(\theta)$ and $R_{\theta} = - \rho sin ( \theta )$.  The up-down symmetry means that $z_{k}^{s} = r_{k}^{c} = 0 $ and the simple shaping means that we only have coupling between deformations and radial fields whose poloidal mode number differs by 1.  The full relationship is
\begin{eqnarray}
B_{\rho m}^{s} sin ( \alpha ) + B_{\rho m}^{c} cos ( \alpha ) = \nonumber\\
(n - (m+1)/q) \frac{\rho }{R_{0}} ( \gamma^{1}_{+1} sin( \alpha ) + \gamma^{2}_{+1} cos ( \alpha ) \\
B_{\rho m-1}^{s} sin ( \alpha_{-1} ) + B_{\rho m-1}^{c} cos ( \alpha_{-1} ) = \nonumber\\
 (n - (m)/q) \frac{\rho }{R_{0}} ( \gamma^{1}_{0} sin( \alpha_{-1} ) + \gamma^{2}_{0} cos ( \alpha_{-1} ) \\
B_{\rho m-2}^{s} sin ( \alpha_{-2} ) + B_{\rho m-2}^{c} cos ( \alpha_{-2} )  = \nonumber\\
 (n - (m-1)/q) \frac{\rho }{R_{0}} ( \gamma^{1}_{-1} sin( \alpha_{-2} ) + \gamma^{2}_{-1} cos ( \alpha_{-2} ) .
\end{eqnarray}
It is important to note that near the $q=m/n$ surface, the largest deformation is produced by the radial magnetic perturbation with helicity $(m-1)/n$.  To lowest order it is not the resonant radial magnetic perturbation that is the most important in terms of producing deformations.  The point is that the deformations are largely governed by non-resonant radial fields and that these deserve more attention.  

Realistic 2D shaping and finite aspect ratio increase the amount of coupling between the perturbation spectrum and the deformation spectrum.  For example, elongation modifies the $k=1$ component of $Z_{\Theta}$ which couples the $(m+1)/n$ field to the resonant deformation and proportionatly weakens the coupling of the $(m-1)/n$ field to the resonant deformation.  Triangularity, finite aspect ratio, and up-down asymmetry also have distinct effects on the coupling.  Given the shape of the axisymmetric flux surface, Equation \ref{fullcoupling} provides a guide for how to tailor the spectral content of 3D fields to control the magnitude of 3D deformations.  

We note in closing that recent modeling has demonstrated the importance of the plasma response in determining the 3D fields present in the plasma \cite{ferraro_nf13}.  Global, 2-fluid equilibrium calculations have found that the deformation-inducing kink response is due to plasma amplification of non-resonant 3D fields, consistent with our calculations \cite{wingen}.  ELM suppression at DIII-D with $n=2$ fields was achieved when the spectral content of the RMPs was chosen to be non-resonant near the pedestal top \cite{lanctot}, consistent with our hypothesis that deformations due to the kink response can enhance turbulent transport to allow for ELM suppression.  

In summary, we have analysed the geometric consequences of RMP-induced 3D deformations of flux surfaces in Tokamaks (if and when they remain intact).  Near a rational surface, only resonant or near-resonant deformations can be driven to significant amplitudes by magnetic perturbations of experimentally relevant magnitude.  We have derived a coupling matrix which determines which parts of the 3D field spectrum are able to couple to these deformations, highlighting the importance of 2D shaping and the non-resonant parts of the 3D field spectrum.  We have shown how these deformations can strongly modify the magnetic curvature and local magnetic shear due to their fine scale spatial structure (i.e. large poloidal mode number).  Any instabilities which tap into the pressure/curvature coupling will be affected by this, including microinstabilities which drive turbulent transport as well as the macroscopic Peeling-Ballooning modes which are thought to drive ELMs.  

These results offer a novel theoretical framework for interpreting results of RMP experiments, and a more detailed analysis of experimental results with these ideas in mind will be the focus of future studies.

\begin{center}
\bf{ACKNOWLEDGMENTS}
\end{center}

T.B. would like to thank Per Helander for useful discussions.   This research was supported by the U.S. Department of Energy under grant
nos.\ DE-FG02-99ER54546 and DE-FG02-86ER53218.  This project has also received funding from the European Union's Horizon
2020 research and innovation programme under grant agreement number
633053. The views and opinions expressed herein do not necessarily
reflect those of the European Commission.

\end{document}